\documentstyle[aps,12pt,manuscript]{revtex}
\begin{document}
\preprint{INJE-TP-98-3/hep-th9803227}

\title{Probing the BTZ black hole with test fields}

\author{ H.W. Lee, N.J. Kim and Y. S. Myung}
\address{Department of Physics, Inje University, Kimhae 621-749, Korea} 

\maketitle

\begin{abstract}
We introduce  a set of scalar fields as  test fields to 
study the dynamical behaviors of 
the BTZ(Banados-Teitelboim-Zanelli) black hole.
These include minimally coupled scalar, conformally coupled scalar, 
dilaton, and tachyon. To calculate the decay rate of the BTZ black hole, 
we consider both the Dirichlet boundary condition at spatial infinity 
and the stability condition. It turns out that 
the tachyon may be a relevant field to get information of the 
BTZ black hole.
\end{abstract}

\newpage
Recently anti-de Sitter spacetime(AdS) has attracted much interest. It 
appears that the conjecture relating 
the string theory on AdS to conformal field theory(CFT) on the boundary 
may resolve many problems in black hole physics\cite{Mal97}.
Sfetsos and Skenderis showed that 
4d black hole(5d black hole) correspond to U-dual to 
BTZ$\times S^2$(BTZ$\times S^3$)\cite{Sfe97}. They calculated the entropies 
of non-extremal 4d and 5d black holes by applying Carlip's 
approach to the BTZ black hole. The BTZ black hole(locally, 
(2+1)-dimensional anti-de Sitter spacetime : AdS$_3$)
has no curvature singularity\cite{Ban92} 
and is considered as a prototype for the general CFT/AdS correspondence
\cite{Mal97}.
This is actually an exact solution of string theory
\cite{Hor93}  
and there is an exact CFT with it on the 
boundary. Carlip has shown that the physical boundary 
degrees of freedom account for the Bekenstein-Hawking 
entropy of the BTZ black hole correctly\cite{Car95}.

In this letter we investigate the 
dynamical behavior of 
the BTZ black hole with test scalar fields. 
Apart from counting the microstates 
of black holes, the dynamical behavior is also an important issue
\cite{Dha96,Cal97}. This is 
so because the greybody factor(absorption cross-section) for 
the black hole arises as a consequence 
of scattering of test field off the gravitational 
potential barrier surrounding the horizon.  
That is, this is an effect of spacetime curvature. Together with the 
Bekenstein-Hawking entropy, this seems to be the strong hint of a deep 
and mysterious connection between curvature and statistical mechanics.
Birmingham {\it et al.} showed that the greybody 
factor for the BTZ black hole has the 
same form as the one for 5d black hole in the dilute gas approximation
\cite{Bir97}. 
However a minimally coupled scalar was used for this end and 
the boundary condition at spatial infinity is not required for calculation. 
It is found that the dilaton(fixed scalar) plays an 
important role in probing the dynamical 
behaviors, when one does not impose any boundary condition at the 
infinity\cite{Lee9803}.  Also a conformally coupled scalar may be used 
for studying the BTZ black hole\cite{Lif94}. 
But it turns out that this is not appropriate for describing the particle 
creation by the BTZ black hole\cite{Kim97}.  As a result, it is 
very important to inquire which one is suitable for studying 
the dynamical behavior of the BTZ black hole among a set of scalar fields.

We start with the BTZ black hole spacetime as
\begin{eqnarray}
&& \bar g_{\mu\nu} =
 \left(  \begin{array}{ccc} - (M - {r^2 / l^2}) & -{J / 2} & 0  \\
                             -{J / 2} & r^2 & 0  \\
    0 & 0 & f^{-2}
         \end{array}
 \right)
\label{bck_metric}
\end{eqnarray}
with $f^2 =r^2 / l^2 -M +J^2 / 4 r^2$.
The metric $\bar g_{\mu\nu}$ is singular at $r=r_{\pm}$,
\begin{equation}
r_{\pm} = {{Ml^2} \over 2} \left \{ 1 \pm \left [ 
   1 - \left ( {J \over Ml} \right )^2 \right ]^{1/2} \right \}
\label{horizon}
\end{equation}
with $M=(r_+^2 + r_-^2) / l^2, J=2 r_+r_- / l$.
Here $r_+(r_-)$ correspond to the outer(inner) horizon. 
For convenience, we list the Hawking temperature $T_H$, the area of 
horizon ${\cal A}_H$, and the angular velocity at the horizon
$\Omega_H$ as
\begin{equation}
T_H = (r_+^2 - r_-^2) / 2 \pi l^2 r_+,
~~{\cal A}_H = 2 \pi r_+,
~~\Omega_H = J / 2 r_+^2.
\label{temp}
\end{equation}
To study the propagation specifically, we introduce 
the small test field($\Psi$)
around the background solution 
(\ref{bck_metric}).
In three dimensions we have no 
propagating degrees of freedom for graviton $h_{\mu\nu}$\cite{Lee9803} 
and vector fields are dual to scalar fields\cite{Bak97}.  
Also the two-form field $B_{\mu\nu}$ has no physical 
degrees of freedom for d=3.
Hence the physical degrees of freedom in the BTZ black hole turn out to be 
a set of scalar fields which satisfy 
\begin{equation}
\bar \nabla^2 \Psi + { s \over l^2} \Psi = 0,
\label{eq_Psi}
\end{equation} 
where $s$ is a new coupling constant. 
$s=0, 3/4, -8$ cases correspond to minimally coupled scalar, 
conformally coupled scalar, and dilaton, respectively.
Considering the $t$ and $x$-translational symmetries of the background 
spacetime (\ref{bck_metric}),  
we can decompose $\Psi$ into frequency modes in these 
variables \cite{Gre93} as
\begin{eqnarray}
\Psi(t,x,r)&&=e^{-i \omega t} e^{i \mu x} \tilde\Psi(r), ~ \mu \in Z.
\label{ptr_scalar}
\end{eqnarray}
Together (\ref{eq_Psi}) with (\ref{bck_metric}) leads to
\begin{eqnarray}
\left [ f^2 \partial_r^2 
+ \left\{ {1 \over r} (\partial_r rf^2) \right\} \partial_r
-{{J \mu \omega} \over {r^2 f^2}} 
+{\omega^2 \over f^2} 
+{{M-{r^2 \over l^2}} \over r^2 f^2} \mu^2 
\right ] \tilde \Psi
+{s \over l^2}\tilde \Psi 
=0.
\label{eq_decoupled}
\end{eqnarray}

Let us calculate the absorption cross-section 
to study the dynamical behavior of 
the BTZ black hole.  
First we note that the spatial infinity of the BTZ black hole is 
timelike(AdS$_3$ is not globally hyperbolic), 
so that information may enter or exit from the boundary at 
infinity.  Thus one impose the boundary condition at infinity to 
obtain the sensible results\cite{Lif94}.  
Here we require the Dirichlet condition as
\begin{equation}
\lim_{r\to \infty} \sqrt{r} \tilde \Psi_\infty =0.
\label{Dirichlet}
\end{equation}
Second, since it is hard to find a 
solution to (\ref{eq_decoupled}) directly, 
we use a matching procedure. 
The spacetime is divided into two regions: the near region ($r \sim r_+$) 
and far region ($r \to \infty$)\cite{Dha96,Cal97}.  
We now study each region in turn.
For the far region($r \to \infty$), (\ref{eq_decoupled}) 
becomes
\begin{equation}
\tilde \Psi_\infty'' + {3 \over r} \tilde \Psi_\infty' 
+ {s \over r^2} \tilde \Psi_\infty=0.
\label{eq_far}
\end{equation}
Here one finds the far region solution
\begin{equation}
\tilde \Psi_{\rm far}(r) = 
{1 \over r} \left ( A_1r^{-\sqrt{1-s} } 
                  + A_2r^{+\sqrt{1-s} } \right ).
\label{sol_far0}
\end{equation}

In order to obtain the near region behavior,
we introduce the variable 
$z={{r^2 - r_+^2} \over {r^2 - r_-^2}},~~0 \le z \le 1$.
Then (\ref{eq_decoupled}) becomes
\begin{equation}
z(1-z) {{d^2 \tilde \Psi} \over dz^2}
+(1-z) {{d \tilde \Psi} \over d z}
+\left ( {A_1 \over z} +{{s/4} \over 1-z} +B_1 \right ) \tilde \Psi
=0,
\label{eq_hyper0}
\end{equation}
where 
$A_1 = 
\left ( {{\omega - \mu \Omega_H} \over 4 \pi T_H} \right )^2, 
B_1 = - {r_-^2 \over r_+^2}
\left ({{\omega - \mu \Omega_H r_+^2 / r_-^2} \over 
4 \pi T_H} \right )^2$.
The solution for (\ref{eq_hyper0}) is given by the hypergeometric functions 
\begin{eqnarray}
\tilde \Psi_{\rm near}(z) &=&
C_1 z^{-i \sqrt{A_1}} (1 -z )^{(1 - \sqrt{1-s})/2} F(a,b,c;z)
\nonumber \\
&&~~~~~~~~+C_2 z^{i \sqrt{A_1}} (1 -z )^{(1 - \sqrt{1-s})/2} F(b-c+1,a-c+1,2-c;z),
\label{sol_hyper0}
\end{eqnarray}
where
\begin{eqnarray}
a&=& \sqrt{B_1} - i \sqrt{A_1} +(1 -  \sqrt{1-s} )/2,
\nonumber \\
b&=& - \sqrt{B_1} - i \sqrt{A_1} + (1 -  \sqrt{1-s} )/2,
\nonumber \\
c&=& 1 - 2 i \sqrt{A_1}.
\label{abc} 
\end{eqnarray}
and $C_1$ and $C_2$ are to-be-determined constants.
At the near horizon($r\sim r_+, z \sim 0$), (\ref{sol_hyper0})
becomes
\begin{eqnarray}
\tilde \Psi_{\rm near}(0) &\simeq& C_1 z^{-i \sqrt{A_1}} + C_2 z^{i \sqrt{A_1}}
\nonumber \\
&=&C_1 \left ( { 2 r_+ \over {r_+^2- r_-^2}} \right ) ^{-i \sqrt{A_1}}
           e^{-i \sqrt{A_1} \ln(r-r_+)}
+C_2 \left ( { 2 r_+ \over {r_+^2- r_-^2}} \right ) ^{i \sqrt{A_1}}
           e^{i \sqrt{A_1} \ln(r-r_+)}.
\label{sol_z0}
\end{eqnarray}
Considering an ingoing mode at horizon, we have $C_2=0$. 
Hence the near region solution is
\begin{eqnarray}
\tilde \Psi_{\rm near}(z) &=&
C_1 z^{-i \sqrt{A_1}} (1 -z )^{(1 - \sqrt{1-s})/2} F(a,b,c;z).
\label{sol_near0}
\end{eqnarray}
To use matching procedure we need to know the behaviour of 
 (\ref{sol_near0}) near $z=1$. This can be obtained by 
the transformation rule ($z\to 1-z$) for hypergeometric function\cite{Abr66}.  
Using this rule we can obtain
\begin{eqnarray}
\tilde \Psi_{n\to f}(z) &=&
{1 \over r} \left ( 
C_1 E_1r^{ - \sqrt{1-s}}  
+C_1 E_2r^{+ \sqrt{1-s}} \right ) ,
\label{far_z0}
\end{eqnarray}
where
\begin{eqnarray}
E_1 &=& 
{{ \Gamma(1 -2 i \sqrt{A_1}) \Gamma( -\sqrt{1-s})
    (r_+^2-r_-^2)^{(1+\sqrt{1-s})/2} }\over
 {\Gamma({1 \over 2} + \sqrt{B_1} -i  \sqrt{A_1} - {\sqrt{1-s} \over 2}))
  \Gamma({1 \over 2} - \sqrt{B_1} -i  \sqrt{A_1} - {\sqrt{1-s} \over 2}))}},
\label{E1} \\
E_2 &=&
{{ \Gamma(1 -2 i \sqrt{A_1}) \Gamma(\sqrt{1-s})
    (r_+^2-r_-^2)^{(1-\sqrt{1-s})/2} }\over
  {\Gamma({1 \over 2} - \sqrt{B_1} -i \sqrt{A_1} + {\sqrt{1-s} \over 2}))
   \Gamma({1 \over 2} + \sqrt{B_1} -i \sqrt{A_1} + {\sqrt{1-s} \over 2}))}}.
\label{E2}
\end{eqnarray}

In matching (\ref{sol_far0}) with (\ref{far_z0}), one needs to 
classify the next steps according to the values of $s$.

(i) $s>1$ case

\noindent
In this case, (\ref{sol_far0}) becomes
\begin{equation}
\tilde \Psi_{\rm far}(r) = 
{1 \over r} \left ( A_{\rm in}e^{-i\sqrt{s-1} \ln r} 
                  + A_{\rm out}e^{+i\sqrt{s-1} \ln r} \right )
\label{sol_far1}
\end{equation}
with two unknown coefficients $A_{\rm in}$ and $A_{\rm out}$.
The first(second) term correspond to ingoing(outgoing) waves at the far region.
This can be confirmed by calculating the corresponding 
flux as ${\cal F}_{\rm in} = - 4 \pi \sqrt{s-1} | A_{\rm in} |^2$
(${\cal F}_{\rm out} =  4 \pi \sqrt{s-1} | A_{\rm out} |^2$).
Comparing this with (\ref{far_z0}) leads to 
$A_{\rm in} = C_1 E_1$ and $A_{\rm out} = C_1 E_2$.
The absorption coefficient is calculated as 
\begin{equation}
{\cal A} = 
1 - \left | {A_{\rm out} \over A_{\rm in} }\right |^2 =
{{(e^{2\sqrt{s-1} \pi} -1 )(e^{\omega' / T_H} -1)} 
\over
 {\left(e^{{\omega'_R / 2 T_R} + \sqrt{s-1} \pi} +1\right) 
  \left(e^{{\omega'_L / 2 T_L} + \sqrt{s-1} \pi} +1\right)}},
\label{cross} 
\end{equation} 
where $\omega' = \omega - \mu \Omega_H, \omega'_{L/R} = \omega \mp \mu \Omega_H r_+ /r_-$ and 
left/right temperatures are defined by
\begin{equation}
{1 \over T_{L/R}} = {1 \over T_H} \left ( 1 \pm {r_- \over r_+} \right ).
\label{temperature}
\end{equation}
The absorption corss-section is given by 
$\sigma_{\rm abs} = {\cal A} / \omega$ in three dimensions.
Finally the decay rate is given as 
\begin{equation}
\Gamma_{\rm fixed} = { \sigma_{\rm abs} \over {e^{\omega' \over T_H} -1 }}
= {1 \over \omega} {{e^{2 \sqrt{s-1} \pi}-1} \over 
 {\left(e^{{\omega'_R / 2 T_R} + \sqrt{s-1} \pi} +1\right) 
  \left(e^{{\omega'_L / 2 T_L} + \sqrt{s-1} \pi} +1\right)}}.
\label{decay_rate}
\end{equation}

(ii) $s=1$ case

\noindent
Here the far region solution is 
\begin{equation}
\tilde \Psi_{\rm far}(r) = 
{1 \over r} \left ( A_1
                  + A_2 \ln r \right ).
\label{sol_far20}
\end{equation} 
For our purpose this can be rearragned to give
\begin{equation}
\tilde \Psi_{\rm far}(r) = 
{A_{\rm in} \over r} \left ( 1 - i \ln r \right )
+{A_{\rm out} \over r} \left ( 1 + i \ln r \right ).
\label{sol_far2}
\end{equation} 
The first term give us the ingoing flux (
${\cal F}_{\rm in} = -4 \pi |A_{\rm in}|^2$ ) and the second is 
outgoing flux (${\cal F}_{\rm out} = -4 \pi |A_{\rm out}|^2$).
We note that 
the transformation rule in deriving (\ref{far_z0}) 
is valid only for $c \ne a+b \pm m$ with integer $m$.  
Considering (\ref{abc}) with $s=1$, we have $c=a+b$. 
In this case we should use a different rule for $z\to 1-z$\cite{Abr66}. 
Using this we obtain
\begin{eqnarray}
\tilde \Psi_{n\to f}(z) &=&
{1 \over r} \left ( 
C_1 D_1  
-C_1 D_2 \ln r \right ) ,
\label{far_z2}
\end{eqnarray}
where
\begin{eqnarray}
D_1 &=& 
{{ \Gamma(1 -2 i \sqrt{A_1}) 
    (r_+^2-r_-^2)^{1/2} }\over
 {\Gamma({1 \over 2} + \sqrt{B_1} -i  \sqrt{A_1} )
  \Gamma({1 \over 2} - \sqrt{B_1} -i  \sqrt{A_1} )}}
[ 2 \psi(1)-\psi(a)-\psi(b)-\ln(r_+^2-r_-^2)]
,
\label{D1} \\
D_2 &=&
-2 {{ \Gamma(1 -2 i \sqrt{A_1}) 
    (r_+^2-r_-^2)^{1/2} }\over
 {\Gamma({1 \over 2} + \sqrt{B_1} -i  \sqrt{A_1} )
  \Gamma({1 \over 2} - \sqrt{B_1} -i  \sqrt{A_1} )}}.
\label{D2}
\end{eqnarray}
Here $\psi$ is the digamma function. 
Comparing (\ref{sol_far2}) and (\ref{far_z2}), one finds 
$A_{\rm in}=C_1(D_1 -i D_2 )/2$ and $A_{\rm out} = C_1 (D_1 + i D_2)/2$.
The absorption coefficient is calculated as
\begin{equation}
{\cal A} = 
1 - \left | {A_{\rm out} \over A_{\rm in} }\right |^2 =
{{8 \Im\left [2 \psi(1)-\psi(a)-\psi(b)-\ln\Delta_-\right ]} 
\over
 {\left \vert 2 \psi(1)-\psi(a)-\psi(b)-\ln\Delta_--2 i\right \vert ^2}}
\label{cross2} 
\end{equation} 
with $\Delta_-=(r_+^2-r_-^2)$.
Here $\Im[\cdots]$ means the imaginary part of its argument.
The tachyon with $s > 0$ may come from the 
low energy string effective action\cite{Ban91}. 
This may be adjusted to satisfy (\ref{eq_Psi}) with $0 < s \le 2$, when 
one chooses a suitable convention for the BTZ black hole\cite{Gho94}. 
In this sense the tachyon may play a role in deriving information for 
the BTZ black hole.

(iii) ${3 \over 4} < s < 1$ case

\noindent
In this regime, (\ref{sol_far0}) takes the form
\begin{equation}
\tilde \Psi_{\rm far}(r) = 
{A_{\rm in} \over r} \left ( r^{\sqrt{1-s}} + ir^{-\sqrt{1-s}}  \right )
+{A_{\rm out} \over r} \left ( r^{\sqrt{1-s}} - ir^{-\sqrt{1-s} } \right ).
\label{sol_far3}
\end{equation}
The first(second) term correspond to ingoing flux 
$ = - 8 \pi \sqrt{1-s} |A_{\rm in}|^2$ (outgoing flux
$ =  8 \pi \sqrt{1-s} |A_{\rm out}|^2$).  From 
(\ref{sol_far3}) and (\ref{sol_z0}) we find 
$A_{\rm in} = -i C_1 (E_1 + i E_2) /2 $
$A_{\rm out} = i C_1 (E_1 - i E_2) /2 $. The absorption coefficient is
\begin{equation}
{\cal A} = 
1 - \left | {A_{\rm out} \over A_{\rm in} }\right |^2 =
1-
{{|E_1 - i E_2|^2} 
\over
 {|E_1 + i E_2|^2}},
\label{cross3} 
\end{equation} 
where $E_1$ and $E_2$ are given by (\ref{E1}) and (\ref{E2}).
In the extremal limit($r_+=r_-$) one finds the total reflection
(${\cal A} = 0$).

(iv) $0 < s \le {3 \over 4}$ case

\noindent
Equation (\ref{sol_far0}) can be rewritten as
\begin{equation}
\tilde \Psi_{\rm far}(r) = 
{1 \over r} \left [ \left ( A_{\rm in} + A_{\rm out} \right ) e^{\sqrt{1-s}} 
+i \left ( A_{\rm in} - A_{\rm out} \right ) e^{-\sqrt{1-s}} \right ] .
\label{sol_far4}
\end{equation}
If one requires the boundary condition, then one finds 
$A_{\rm in} = - A_{\rm out}$.  In this case the total flux
(${\cal F}_{\rm in} + {\cal F}_{\rm out}$) is zero, and 
so one finds the total reflection.
Further (\ref{sol_far4}) with $A_{\rm in} =- A_{\rm out}$ 
cannot be matched with (\ref{far_z0}).  
This means that 
the field with $s \le 3/4$ is not 
useful for probing the dynamical 
behaviour of BTZ black hole.  We note that a conformally coupled 
scalar takes the value $s=3/4$\cite{Lif94}.  As a result, this field 
seems to be not 
appropriate for studying the particle creation by the BTZ black hole.
Also Kim {\it et al.} pointed out that the conformally coupled
scalar is not useful for investigating the Hawking radiation
of the BTZ black hole. They showed that if one uses a conformally coupled
matter in the conformally-flat spacetime, there is no particle
creation in the BTZ black hole. This supports that our 
result for a conformally coupled scalar is correct.

(v) $s=0$ case

\noindent
In this case, (\ref{sol_far0}) takes the form
\begin{equation}
\tilde \Psi_{\rm far}(r) = 
A_1   + {A_2 \over r^2} . 
\label{sol_far5}
\end{equation}
Using the 
transformation rule for $c=a+b+1$\cite{Abr66}, we obtain 
the asymptotic form 
\begin{eqnarray}
\tilde \Psi_{n\to f}(z) &\simeq&
C_1 F_1 + {C_1 F_2 \over r^2} - {{C_1 F_3 \ln r} \over r^2}, 
\label{far_z5}
\end{eqnarray}
where
\begin{eqnarray}
F_1 &=& {\Gamma(a+b+1) \over {\Gamma(a+1)\Gamma(b+1)}},
\nonumber \\
F_2 &=& \Delta_-[\ln\Delta_- +\psi(a+1)+\psi(b+1)-\psi(1)-\psi(2)]
ab F_1
\nonumber \\
F_3 &=& \Delta_- ab F_1.
\nonumber \\
\end{eqnarray}
Here $ab=- l^2 (\omega^2 -\mu^2)/2\Delta_-$.
If $\mu=0$ and $\omega \ll 1/l$, $F_2$ and $F_3$ is small compared to
$F_1$. Thus we find $A_1=C_1F_1$ and $A_2 \simeq 0$. 
However this does not satisfy the Dirichlet boundary condition unless 
$A_1=C_1F_1=0$. That is, we have no information for the BTZ black hole 
with a minimally coupled scalar($s=0$).  
Birmingham {\it et al.} calculated the decay rate of the BTZ black 
hole with this scalar\cite{Bir97}, 
\begin{equation}
\Gamma_{\rm min} 
= { \sigma_{\rm abs}^{\rm min} \over {e^{\omega \over T_H} -1 }}
= {{\pi^2 l^2 \omega} \over 
 {(e^{\omega \over 2 T_R} -1) 
  (e^{\omega \over 2 T_L} -1)}}.
\label{decay_min}
\end{equation}
However they did not ask the boundary condition at spatial 
infinity. If one requires (\ref{Dirichlet}) 
it is hard to calculate the decay rate for the BTZ black hole.

(vi) $s<0$ case

\noindent
Here, (\ref{sol_far0}) can be rewritten as
\begin{equation}
\tilde \Psi_{\rm far}(r) = 
{1 \over r} \left [ \left ( A_{\rm in} + A_{\rm out} \right ) r^{\sqrt{1-s}}
+i\left ( A_{\rm in} - A_{\rm out} \right ) r^{-\sqrt{1-s}} \right ].
\label{sol_far6}
\end{equation}
Considering the boundary condition, then one finds 
$A_{\rm in} = - A_{\rm out}$.  In this case the total flux
(${\cal F}_{\rm in} + {\cal F}_{\rm out}$) is zero, 
and this means the total reflection.
Further (\ref{sol_far6}) with $A_{\rm in} = -A_{\rm out}$ 
cannot be matched with (\ref{far_z0}).  
This implies that 
the field with $s < 0$ may not be 
useful for probing the dynamical behaviour of BTZ black hole.  
The dilaton and massive scalar field which satisfies 
$\nabla^2 \Psi - m^2 \Psi=0$, belong to this case\cite{Lee9803,Ich95}.

In conclusion, we introduce a set of scalar fields to study 
the Hawking radiation of the BTZ black hole. These are minimally 
coupled scalar, conformally coupled scalar, and dilaton including 
tachyon and massive scalar field. 
If one does not require the boundary condition at spatial infinity, 
the dilaton can be regarded as the 
best field to get information about the 
BTZ black hole.  
As was shown in the $s<0$ case, it turns out that the dilaton with 
$s=-8$ is not appropriate for probing the BTZ black hole.
However, the tachyon may be emerged 
as a candidate for inquiring the BTZ black hole. 
The tachyon is defined as a field with $m^2 <0$ and thus causes the 
difficulty in the Minkowski spacetime. On the other hand, the tachyon 
may play an important role in the anti-de Sitter spacetime.
Because of the 
boundary condition at spatial infinity, the kinetic energy of a 
scalar field in AdS$_3$ cannot vanish. Thus the stability  of the 
BTZ black hole requires 
not that $m^2$ should be positive ($s<0$) but $m^2$ should be 
no smaller than a certain negative lower bound\cite{Wit98}. For d-dimensional 
AdS case, this is approximately given by $m^2 \ge -(d-1)^2/4$. 
Then test fields should satisfy the stability condition : 
$m^2 \ge -1 ( s \le 1 )$. 

Hence if we take both the stability
($s \le 1$) of the 
BTZ black hole and the boundary condition at spatial infinity
($s > 3/4$) seriously, the relevant field will be the 
tachyon with $3/4 < s \le 1$. Also the dilaton with $s=-8$ may be 
excluded from our consideration.
In this sense, the remaining fields (minimally coupled 
scalar($s=0$), conformally coupled scalar($s=3/4$), and massive scalar 
field($s < 0$)) are not suitable for our purposes.

\section*{Acknowledgement}
This work was supported in part by the Basic Science Research Institute 
Program, Minstry of Education, Project NOs. BSRI-97-2413 and 
BSRI-97-2441.

\end{document}